\documentclass[conference,10pt,twocolumn]{IEEEtran}
\usepackage{cite}
\usepackage[T1]{fontenc}
\usepackage{graphicx}
\usepackage{amsmath}
\usepackage{amssymb}
\usepackage{multirow}

\renewcommand{\eqref}[1]{(\ref{#1})}
\newcommand{\secref}[1]{\mbox{Section~\ref{#1}}}

\newcommand{\figref}[1]{\mbox{Fig.~\ref{#1}}}

\newcommand{\NIT}{\ensuremath{I}}
\newcommand{\IEmap}[1]{f\!\left(#1\right)} 
\newcommand{\EImap}[1]{f^{-1}\!\left(#1\right)} 

\renewcommand{\Pr}[1]{\ensuremath{\mathrm{P}\!\left[#1\right]}} 
\newcommand{\PrT}[1]{\ensuremath{\mathrm{P}\left[#1\right]}} 

\DeclareMathOperator*{\argmin}{arg\;min}

\DeclareMathOperator*{\define}{\triangleq} 

\newcommand{\setA}{\ensuremath{\mathcal{A}}}

\newcommand{\setO}{\ensuremath{\mathcal{O}}}

\newcommand{\setX}{\ensuremath{\mathcal{X}}}

\newcommand{\setZ}{\ensuremath{\mathcal{Z}}}

\newcommand{\bmb}{\ensuremath{\mathbf{b}}}

\newcommand{\bmn}{\ensuremath{\mathbf{n}}}

\newcommand{\bms}{\ensuremath{\mathbf{s}}}

\newcommand{\bmx}{\ensuremath{\mathbf{x}}}
\newcommand{\bmy}{\ensuremath{\mathbf{y}}}

\newcommand{\bA}{\ensuremath{\mathbf{A}}}

\newcommand{\bH}{\ensuremath{\mathbf{H}}}
\newcommand{\bI}{\ensuremath{\mathbf{I}}}

\newcommand{\bQ}{\ensuremath{\mathbf{Q}}}
\newcommand{\bR}{\ensuremath{\mathbf{R}}}

\newcommand{\MT}{\ensuremath{M_{T}}}
\newcommand{\MR}{\ensuremath{M_{R}}}

\newcommand{\const}{\ensuremath{\setO}}
\newcommand{\mimoconst}{\ensuremath{\setO^{\MT}}}
\newcommand{\modorder}{\ensuremath{Q}}

\def\addnotation #1: #2{\parbox[t]{1.25cm}{$#1$\dotfill} \parbox[t]{3.50in}{#2} \vspace{0.0625cm} \\}
\def\addsymbol #1: #2{\parbox[t]{2cm}{$#1$} \parbox[t]{3in}{#2} \vspace{0.0625cm} \\}
\def\addacronym #1: #2{\parbox[t]{2cm}{\bf{#1}\dotfill} \parbox[t]{2.8in}{#2} \vspace{0.0625cm} \\}

\title{Soft-Input Soft-Output Sphere Decoding}

\author{
\IEEEauthorblockN{Christoph~Studer}
\IEEEauthorblockA{Integrated Systems Laboratory \\
ETH Zurich, 8092 Zurich, Switzerland \\
Email: studer@iis.ee.ethz.ch
\vspace{-0.29cm}}
\and
\IEEEauthorblockN{Helmut~B\"olcskei}
\IEEEauthorblockA{Communication Technology Laboratory \\
ETH Zurich, 8092 Zurich, Switzerland \\
Email: boelcskei@nari.ee.ethz.ch
\vspace{-0.29cm}}
}

\begin{document}

\maketitle

\begin{abstract}
  Soft-input soft-output (SISO) detection algorithms form the basis
  for iterative decoding. The associated computational complexity
  often poses significant challenges for practical receiver
  implementations, in particular in the context of multiple-input
  multiple-output wireless systems. In this paper, we present a
  low-complexity SISO sphere decoder which is based on the single tree
  search paradigm, proposed originally for soft-output detection in
  Studer~\emph{et al.}, \emph{IEEE J-SAC}, 2008. The algorithm
  incorporates clipping of the extrinsic log-likelihood ratios in the
  tree search, which not only results in significant complexity
  savings, but also allows to cover a large performance/complexity
  trade-off region by adjusting a single parameter.
\end{abstract}


\section{Introduction}

Soft-input soft-output (SISO) detection in multiple-input
multiple-output (MIMO) systems constitutes the basis for iterative
decoding, which, in general, achieves significantly better performance
than decoding based on hard-output or soft-output-only detection
algorithms. Unfortunately, this performance gain comes at the cost of
a significant, often prohibitive (in terms of practical
implementation), increase in terms of computational complexity.

Implementing different algorithms, each optimized for a maximum
allowed detection effort or for a particular system configuration,
would entail considerable circuit complexity. A practical SISO
detector for MIMO systems should therefore not only exhibit low
computational complexity but also cover a wide range of easily
adjustable performance/complexity trade-offs.

The single tree search (STS) soft-output sphere decoder (SD) in
combination with log-likelihood ratio (LLR) clipping~\cite{jsac07} has
been demonstrated to be suitable for VLSI implementation and is
efficiently tunable between max-log optimal soft-output and
low-complexity hard-output detection performance. The STS-SD concept
is therefore a promising basis for efficient SISO detection in MIMO
systems.

\subsubsection*{Contributions}

We describe a SISO STS-SD algorithm that is tunable between max-log
optimal SISO and hard-output maximum a posteriori~(MAP) detection
performance.  To this end, we extend the soft-output STS-SD algorithm
described in~\cite{jsac07} by a max-log optimal a priori information
processing method that significantly reduces the tree-search
complexity compared to, e.g.,~\cite{vikalo02a,jalden05}, and avoids
the computation of transcendental functions. The basic idea of the
proposed approach is to incorporate clipping of the extrinsic LLRs
into the tree search.  This requires that the list administration
concept and the tree pruning criterion proposed for soft-output STS-SD
in~\cite{jsac07} be suitably modified. Simulation results show that
the SISO STS-SD with extrinsic LLR clipping attains close to max-log
optimal SISO performance at remarkably low computational complexity
and, in addition, offers a significantly larger performance/complexity
trade-off region than the soft-output STS-SD in~\cite{jsac07}.

\subsubsection*{Notation}

Matrices are set in boldface capital letters, vectors in boldface
lowercase letters. The superscripts $^T$ and $^H$ stand for transpose
and conjugate transpose, respectively.  We write $A_{i,j}$ for the
entry in the $i$th row and $j$th column of the matrix $\bA$ and $b_i$
for the $i$th entry of the
vector~\mbox{$\bmb=[\,b_1\,\,b_2\,\,\cdots\,\,b_N\,]^T$}. $\bI_{N}$
denotes the $N\times N$ identity matrix. Slightly abusing common
terminology, we call an $N\times M$ matrix $\bA$, where $N\geq M$,
satisfying $\bA^H\bA=\bI_{M}$, unitary. $|\setO|$ denotes the
cardinality of the set $\setO$. The probability of an event~$\setZ$ is
denoted by $\Pr{\setZ}$. $\overline{x}$ is the binary complement
of~$x\in\{+1,-1\}$, i.e., $\overline{x}=-x$.


\section{Soft-Input Soft-Output Sphere Decoding} \label{SISOSD}

Consider a MIMO system with $\MT$ transmit and $\MR\geq\MT$ receive
antennas. The coded bit-stream to be transmitted is mapped to (a
sequence of) \mbox{\MT-dimensional} transmit symbol
vectors~$\bms\in\mimoconst$, where $\const$ stands for the underlying
complex scalar constellation and $|\setO|=2^\modorder$. Each symbol
vector $\bms$ is associated with a label vector $\bmx$ containing $\MT
Q$ binary values chosen from the set~$\{+1,-1\}$ where the null
element (0 in binary logic) of GF(2) corresponds to~$+1$.  The
corresponding bits are denoted by $x_{j,b}$, where the indices $j$ and
$b$ refer to the $b$th bit in the binary label of the $j$th entry of
the symbol vector $\bms=[\,s_1\,\,s_2\,\,\cdots\,\,s_{\MT}\,]^T$.  The
associated complex baseband input-output relation is given by
\begin{align} \label{eq:channelmodel}
  \bmy = \bH\bms+\bmn
\end{align}
where $\bH$ stands for the $\MR\times\MT$ channel matrix, $\bmy$ is
the $\MR$-dimensional received signal vector, and $\bmn$ is an i.i.d.\
circularly symmetric complex Gaussian distributed
\mbox{\MR-dimensional} noise vector with variance $N_o$ per complex
entry.

\subsection{Max-Log LLR Computation as a Tree Search}

\begin{figure}[t]
  \centering
  \includegraphics[width=0.98\columnwidth]{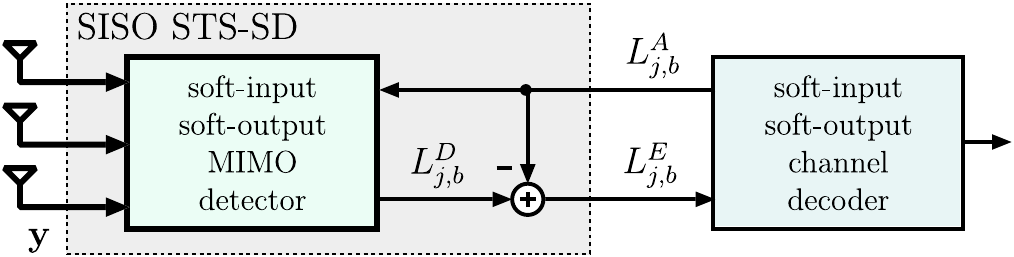}
\vspace{-0.10cm}
\caption{Iterative MIMO decoder~\cite{HB03}. The SISO STS-SD
  (corresponding to the dashed box) directly computes extrinsic
  log-likelihood ratios.}
  \label{fig:decoder}
\vspace{-0.3cm}
\end{figure}

SISO detection for MIMO systems requires computation of the
LLRs~\cite{HB03,Steingrimsson03}
\begin{align} \label{eq:optLLRdef}
L_{j,b} \define \log\Bigg( \frac{\Pr{x_{j,b}=+1|\bmy,\bH}}{\Pr{x_{j,b}=-1|\bmy,\bH}}\Bigg)
\end{align}
for all bits~$j=1,2,\ldots,\MT$, $b=1,2,\ldots,Q$, in the
label~$\bmx$. Transforming \eqref{eq:optLLRdef} into a tree-search
problem and using the sphere decoding algorithm allows efficient
computation of the LLRs~\cite{wang04},\cite{jalden05},\cite{jsac07}.
To this end, the channel matrix $\bH$ is first QR-decomposed according
to $\bH=\bQ\bR$, where the $\MR\times\MT$ matrix $\bQ$ is unitary and
the $\MT\times\MT$ upper-triangular matrix~$\bR$ has real-valued
positive entries on its main diagonal.
Left-multiplying~\eqref{eq:channelmodel} by $\bQ^H$ leads to the
modified input-output relation $\tilde{\bmy} = \bR\bms + \bQ^H\bmn$,
where $\tilde{\bmy}=\bQ^H\bmy$. Noting that $\bQ^H\bmn$ is also
i.i.d.\ circularly symmetric complex Gaussian and using the max-log
approximation leads to the \emph{intrinsic} LLRs~\cite{HB03}
\begin{align} \label{eq:maxlogmapllr} 
  L^D_{j,b}  \define &
  \min_{\bms\in\setX^{(-1)}_{j,b}}\bigg\{
  \frac{1}{N_o}\big\|\tilde{\bmy}-\bR\bms\big\|^2-\log\Pr{\bms}\bigg\}
  \nonumber \\
  - &
  \min_{\bms\in\setX^{(+1)}_{j,b}}\bigg\{\frac{1}{N_o}\|\tilde{\bmy}-\bR\bms\|^2-\log\Pr{\bms}\bigg\}
\end{align}
where $\setX^{(-1)}_{j,b}$ and $\setX^{(+1)}_{j,b}$ are the sets of
symbol vectors that have the bit corresponding to the indices~$j$
and~$b$ equal to~$-1$ and~$+1$, respectively. In the following, we
consider an iterative MIMO decoder as depicted
in~\figref{fig:decoder}. A soft-input soft-output~MIMO detector
computes intrinsic LLRs according to~\eqref{eq:maxlogmapllr} based on
the received signal vector~$\bmy$ and on a priori probabilities in the
form of the \emph{a priori}
LLRs~\mbox{$L^A_{j,b}\define\log\left(\frac{\PrT{x_{j,b}=+1}}{\PrT{x_{j,b}=-1}}\right)$}
and delivers the \emph{extrinsic} LLRs
\begin{align} \label{eq:explicitextrinsicLLRs}
  L^E_{j,b} = L^D_{j,b} - L^A_{j,b}, \qquad \forall\,j,b,
\end{align}
to a subsequent SISO channel decoder.

For each bit, one of the two minima in~\eqref{eq:maxlogmapllr}
corresponds to
\begin{align} \label{eq:mapsymbol}
\lambda^\mathrm{MAP}\define\frac{1}{N_o}\Big\|\tilde{\bmy}-\bR\bms^\mathrm{MAP}\Big\|^2-\log\Pr{\bms^\mathrm{MAP}}
\end{align}
which is associated with the MAP solution of the MIMO detection
problem
\begin{align} \label{eq:MAPsolution}
  \bms^\mathrm{MAP} =
  \argmin_{\bms\in\setO^{\MT}}\bigg\{\frac{1}{N_o}\big\|\tilde{\bmy}-\bR\bms\big\|^2-\log\Pr{\bms}\bigg\}.
\end{align}
The other minimum in \eqref{eq:maxlogmapllr} can be computed as
\begin{align} \label{eq:mapcounterhypometric}
\lambda^{\overline{\mathrm{MAP}}}_{j,b}\define\min_{\bms\in\setX^{\big(\overline{x_{j,b}^\mathrm{MAP}}\big)}_{j,b}} \bigg\{
\frac{1}{N_o}\big\|\tilde{\bmy}-\bR\bms\big\|^2-\log\Pr{\bms}\bigg\}
\end{align}
where the (bit-wise) \emph{counter-hypothesis}
$\overline{x_{j,b}^\mathrm{MAP}}$ to the MAP hypothesis denotes the
binary complement of the $b$th bit in the label of the $j$th entry of
$\bms^\mathrm{MAP}$. With the definitions \eqref{eq:mapsymbol} and
\eqref{eq:mapcounterhypometric}, the intrinsic max-log LLRs
in~\eqref{eq:maxlogmapllr} can be written in compact form as
\begin{align} \label{eq:maxlogmapllrreduced}
L^D_{j,b} = \left\{
\begin{array}{ccc}
 \lambda^{\overline{\mathrm{MAP}}}_{j,b} - \lambda^\mathrm{MAP} \,, &
 x^\mathrm{MAP}_{j,b} = +1 \\
\lambda^\mathrm{MAP} - \lambda^{\overline{\mathrm{MAP}}}_{j,b} \,, & 
x^\mathrm{MAP}_{j,b} = -1.
\end{array}
\right.
\end{align}
We can therefore conclude that efficient max-log optimal soft-input
soft-output MIMO detection reduces to efficiently identifying
$\bms^\mathrm{MAP}$, $\lambda^\mathrm{MAP}$, and
$\lambda^{\overline{\mathrm{MAP}}}_{j,b}$ ($\forall j,b$).

We next define the partial symbol vectors (PSVs)
\mbox{$\bms^{(j)}=[\,s_j\,\,s_{j+1}\,\,\cdots\,\,s_{\MT}\,]^T$} and
note that they can be arranged in a tree that has its root just above
level $j=\MT$ and leaves, on level $j=1$, which correspond to symbol
vectors~$\bms$. The binary-valued label vector associated
with~$\bms^{(j)}$ will be denoted by~$\bmx^{(j)}$. The distances
\begin{align*} 
d(\bms)=\frac{1}{N_o}\big\|\tilde{\bmy}-\bR\bms\big\|^2-\log\Pr{\bms}
\end{align*} 
in~\eqref{eq:mapsymbol} and~\eqref{eq:mapcounterhypometric} can be
computed recursively if the individual symbols~$s_j$
($j=1,2,\ldots,\MT$) are statistically independent, i.e., if
\mbox{$\Pr{\bms}=\prod_{j=1}^{\MT}\Pr{s_j}$}. In this case, we have
\begin{align*}
  d(\bms)=\sum_{j=1}^{\MT}\Bigg(\frac{1}{N_o}\bigg|\tilde{y}_j-\sum_{i=j}^{\MT}R_{j,i}s_i\bigg|^2-\log\Pr{s_j}\Bigg)
\end{align*}
which can be evaluated recursively as $d(\bms)=d_1$, with the partial distances~(PDs)
\begin{align*}
 d_j = d_{j+1} + |e_j|, \quad j=\MT,\MT-1,\ldots,1,
\end{align*}
the initialization~$d_{\MT+1}=0$, and the distance
increments~(DIs)
\begin{align} \label{eq:metricincrements}
|e_j| = \frac{1}{N_o}\Bigg|\tilde{y}_j-\sum_{i=j}^{\MT}R_{j,i}s_i\Bigg|^2 -\log\Pr{s_j}.
\end{align}
Note that the DIs are non-negative since $-\log{\Pr{s_j}}\geq0$. The
dependence of the PDs $d_j$ on the symbol vector $\bms$ is only
through the PSV~$\bms^{(j)}$. Thus, the MAP detection problem and the
computation of the \emph{intrinsic} max-log LLRs have been transformed
into tree-search problems: PSVs and PDs are associated with nodes,
branches correspond to DIs. For brevity, we shall often say ``the
node~$\bms^{(j)}$'' to refer to the node corresponding to the
PSV~$\bms^{(j)}$. We shall furthermore use~$d\big(\bms^{(j)}\big)$
and~$d\big(\bmx^{(j)}\big)$ interchangeably to denote~$d_j$. Each path
from the root node down to a leaf node corresponds to a symbol vector
$\bms\in\setO^{\MT}$. The solution of~\eqref{eq:mapsymbol}
and~\eqref{eq:mapcounterhypometric} corresponds to the leaf associated
with the smallest metric in $\setO^{\MT}$ and
$\setX^{\big(\overline{x_{j,b}^\mathrm{MAP}}\big)}_{j,b}$,
respectively. The SISO STS-SD uses elements of the Schnorr-Euchner SD
with radius reduction~\cite{AEVZ02,burg05}, briefly summarized as
follows: The search in the weighted tree is constrained to nodes which
lie within a radius~$r$ around $\tilde{\bmy}$ and tree traversal is
performed depth-first, visiting the children of a given node in
ascending order of their PDs. A node~$\bms^{(j)}$ with PD~$d_j$ can be
pruned (along with the entire subtree originating from this node)
whenever the pruning criterion~$d_j\geq r^2$ is met. In order to avoid
the problem of choosing a suitable starting radius, we initialize the
algorithm with $r=\infty$ and perform the update~$r^2\gets d(\bms)$
whenever a valid leaf node~$\bms$ has been reached. The complexity
measure employed in the remainder of the paper corresponds to the
number of nodes visited by the decoder including the leaves, but
excluding the root.

\subsection{Tree Search for Statistically Independent Bits}\label{independentbitsearch}

Consider the case where all~$Q$ bits corresponding to a symbol $s_j$
are statistically independent and the MIMO detector obtains a priori
LLRs~$L^A_{j,b}$ ($\forall j,b$) from an external device, e.g., a SISO
channel decoder as depicted in~\figref{fig:decoder}. We then
have~\cite{Hagenauer96}
\begin{align} \label{eq:symbolprob1}
\Pr{s_j} =  \prod_{b=1}^{Q}
\frac{\exp\Big(\frac{1}{2}\big(1+x_{j,b}\big)L^A_{j,b}\Big)}{1+\exp\big(L^A_{j,b}\big)}.
\end{align}
The contribution of the a priori LLRs to the DIs in
\eqref{eq:metricincrements} can be obtained from
\eqref{eq:symbolprob1} as
\begin{align} \label{eq:suboptimalmetric}
-\log\Pr{s_j} = - \sum_{b=1}^{Q} \frac{1}{2} x_{j,b}L^A_{j,b} + \tilde{K}_j
\end{align}
where the constants
\begin{align} \label{eq:maxlognormconstant}
 \tilde{K}_j = \sum_{b=1}^{Q} \bigg( \frac{1}{2} \big|L^A_{j,b}\big| +
 \log\Big(1+\exp\big(\!-|L^A_{j,b}|\big)\Big) \bigg)
\end{align} 
are independent of the binary-valued variables~$x_{j,b}$
and \mbox{$\tilde{K}_j>0$} for $j=1,2,\ldots,\MT$. Because of
$-\log\Pr{s_j}\geq0$, we can trivially infer
from~\eqref{eq:suboptimalmetric} that $-\sum_{b=1}^{Q} \frac{1}{2}
x_{j,b}L^A_{j,b} + \tilde{K}_j \geq 0$.
From~\eqref{eq:maxlogmapllrreduced} it follows that constant terms
(i.e., terms that are independent of the variables~$x_{j,b}$ and hence
of~$\bms$) in~\eqref{eq:mapsymbol} and~\eqref{eq:mapcounterhypometric}
cancel out in the computation of the intrinsic LLRs and can therefore
be neglected. A straightforward method to avoid the
hardware-inefficient task of computing transcendental functions
in~\eqref{eq:maxlognormconstant} is to set~$\tilde{K}_j=0$ in the
computation of~\eqref{eq:suboptimalmetric}. This can, however, lead to
branch metrics that are not necessarily non-negative, which would
inhibit pruning of the search tree.  On the other hand, modifying the
DIs in~\eqref{eq:metricincrements} by setting
\begin{align} \label{eq:BICMmetricincrements}
  |e_j| \define \frac{1}{N_o}\Bigg|\tilde{y}_j-\sum_{i=j}^{\MT}R_{j,i}s_i\Bigg|^2 
 -\sum_{b=1}^{Q} \frac{1}{2}x_{j,b}L^A_{j,b} + K_j
\end{align}
with~$K_j = \sum_{b=1}^{Q} \frac{1}{2} \big|L^A_{j,b}\big|$ also
avoids computing 
trans-cendental 
functions while guaranteeing that,
since \mbox{$-x_{j,b}L^A_{j,b}+\big|L^A_{j,b}\big|\geq0$} ($\forall
j,b$), the so obtained branch metrics are non-negative.  Furthermore,
as~$\tilde{K}_j \geq K_j$, using the modified DIs leads to tighter, but,
thanks to~\eqref{eq:maxlogmapllrreduced}, still max-log optimal tree
pruning, thereby (often significantly) reducing the complexity
compared to that obtained through~\eqref{eq:metricincrements}.

Note that in~\cite[Eq.~9]{baero03}, the prior
term~\eqref{eq:suboptimalmetric} was approximated as
\begin{align*} 
  -\log\Pr{s_{j}}\approx
  \sum_{b=1}^Q\frac{1}{2}\Big(-x_{j,b}L^A_{j,b}+\big|L^A_{j,b}\big|\Big)
\end{align*}
for $\big|L^A_{j,b}\big|>2$ ($b=1,2,\ldots,Q$) which corresponds
exactly to what was done here to arrive
at~\eqref{eq:BICMmetricincrements}. It is important, though, to
realize that using the modified DIs~\eqref{eq:BICMmetricincrements}
does \emph{not} lead to an approximation
of~\eqref{eq:maxlogmapllrreduced}, as the neglected $\log(\cdot)$ term
in~\eqref{eq:maxlognormconstant} does not
affect~\eqref{eq:maxlogmapllrreduced}.

\section{Extrinsic LLR Computation in a Single Tree~Search}
\label{SISOSTS}

Computing the intrinsic max-log LLRs in \eqref{eq:maxlogmapllrreduced}
requires to determine~$\lambda^\mathrm{MAP}$ and the
metrics~$\lambda^{\overline{\mathrm{MAP}}}_{j,b}$ associated with the
counter-hypotheses. For given~$j$ and~$b$ the
metric~$\lambda^{\overline{\mathrm{MAP}}}_{j,b}$ is obtained by
traversing only those parts of the search tree that have leaves
in~$\setX_{j,b}^{\big(x_{j,b}^{\overline{\mathrm{MAP}}}\big)}$. The
quantities~$\lambda^\mathrm{MAP}$ and
$\lambda^{\overline{\mathrm{MAP}}}_{j,b}$ can be computed using the SD
based on the repeated tree search (RTS) approach described
in~\cite{wang04}. The RTS strategy results, however, in redundant
computations as (often significant) parts of the search tree are
revisited during the RTS steps required to
determine~$\lambda^{\overline{\mathrm{MAP}}}_{j,b}$~($\forall j,b$).
Following the STS paradigm described for soft-output sphere decoding
in~\cite{jalden05,jsac07}, we note that \emph{efficient} computation
of $L^D_{j,b}$ ($\forall j,b$) requires that every node in the tree is
visited \emph{at most} once. This can be achieved by searching for the
MAP solution and computing the
metrics~$\lambda^{\overline{\mathrm{MAP}}}_{j,b}$ ($\forall j,b$)
concurrently and ensuring that the subtree emanating from a given node
in the tree is pruned if it can not lead to an update of either
$\lambda^\mathrm{MAP}$ or at least one of
the~$\lambda^{\overline{\mathrm{MAP}}}_{j,b}$. Besides extending the
ideas in~\cite{jsac07} to take into account a priori information, the
main idea underlying the SISO STS-SD presented in this paper is to
\emph{directly} compute the extrinsic LLRs~$L^E_{j,b}$ through a tree
search, rather than computing~$L^D_{j,b}$ first and then
evaluating~\eqref{eq:explicitextrinsicLLRs}.

Due to the large dynamic range of LLRs, fixed-point hardware
implementations need to constrain the magnitude of the LLR value.
Evidently, clipping of the LLR magnitude leads to a degradation in
terms of decoder performance.  It has been noted
in~\cite{Yee05,jsac07} that incorporating LLR clipping into the tree
search is very effective in terms of reducing complexity of max-log
based soft-output sphere decoding. In addition, as demonstrated
in~\cite{jsac07}, LLR clipping, when built into the tree search also
allows to tune the detection algorithm in terms of performance versus
complexity by adjusting the LLR clipping level. In the SISO case, we
are ultimately interested in the \emph{extrinsic} LLRs $L^E_{j,b}$ and
clipping should therefore ensure that $\big|L^E_{j,b}\big|\leq
L_\mathrm{max}$. It is hence sensible to ask whether clipping of the
extrinsic LLRs can be built directly into the tree search. The answer
is in the affirmative and the corresponding solution is described
below.

To prepare the ground for the formulation of the SISO STS-SD, we write
the extrinsic LLRs as
\begin{align} \label{eq:directextrinsicLLRs}
L^E_{j,b} = \left\{
\begin{array}{ccc}
\Lambda^{\overline{\mathrm{MAP}}}_{j,b} - \lambda^\mathrm{MAP} \,, &
 x^\mathrm{MAP}_{j,b} = +1 \\
\lambda^\mathrm{MAP} - \Lambda^{\overline{\mathrm{MAP}}}_{j,b} \,, &
x^\mathrm{MAP}_{j,b} = -1 
\end{array}
\right.
\end{align}
where the quantities
\begin{align} \label{eq:extrinsicmetric}
  \Lambda^{\overline{\mathrm{MAP}}}_{j,b} = \left\{
\begin{array}{ccc}
\lambda^{\overline{\mathrm{MAP}}}_{j,b} - L^A_{j,b} \,, &
 x^\mathrm{MAP}_{j,b}=+1 \\
\lambda^{\overline{\mathrm{MAP}}}_{j,b} + L^A_{j,b} \,, &
 x^\mathrm{MAP}_{j,b}=-1 \\
\end{array}
\right.
\end{align}
will be referred to as the \emph{extrinsic metrics}. For the following
developments it will be convenient to define a function
$\IEmap{\cdot}$ that transforms an intrinsic metric~$\lambda$ with
associated a priori LLR~$L^A$ and binary label~$x$ to an extrinsic
metric~$\Lambda$ according to
\begin{align} \label{eq:intrinsictoextrinsicmapping}
 \Lambda =  \IEmap{\lambda,L^A,x} = \left\{
\begin{array}{cl}
 \lambda - L^A \,, &
 x=+1 \\
 \lambda + L^A \,, &
 x=-1. \\
\end{array}
\right.
\end{align}
With this notation, we can rewrite~\eqref{eq:extrinsicmetric} more
compactly as \mbox{$\Lambda^{\overline{\mathrm{MAP}}}_{j,b} =
  \IEmap{\lambda^{\overline{\mathrm{MAP}}}_{j,b},L^A_{j,b},x^\mathrm{MAP}_{j,b}}$}. The inverse
function of~\eqref{eq:intrinsictoextrinsicmapping} transforms an
extrinsic metric~$\Lambda$ to an intrinsic metric~$\lambda$ and is
defined as
\begin{align} \label{eq:extrinsictointrinsicmapping}
 \lambda =  \EImap{\Lambda,L^A,x} = \left\{
\begin{array}{cl}
 \Lambda + L^A \,, &
 x=+1 \\
 \Lambda - L^A \,, &
 x=-1 .\\
\end{array}
\right.
\end{align}

We emphasize that the tree search algorithm described below produces
the extrinsic LLRs $L^E_{j,b}$ ($\forall j,b$)
in~\eqref{eq:directextrinsicLLRs} rather than the intrinsic ones
in~\eqref{eq:maxlogmapllrreduced}. Consequently, careful modification
of the list administration steps, the pruning criterion, and the LLR
clipping rules of the soft-output algorithm described in~\cite{jsac07}
is needed.

\subsection{List Administration} \label{listadministration}

The main idea of the STS paradigm is to search the subtree originating
from a given node only if the result can lead to an update of either
$\lambda^\mathrm{MAP}$ or of at least one of the
$\Lambda^{\overline{\mathrm{MAP}}}_{j,b}$. To this end, the decoder
needs to maintain a list containing the label of the current MAP
hypothesis~$\bmx^\mathrm{MAP}$, the corresponding
metric~$\lambda^\mathrm{MAP}$, and all $Q\MT$ extrinsic
metrics~$\Lambda^{\overline{\mathrm{MAP}}}_{j,b}$. The algorithm is
initialized with
\mbox{$\lambda^\mathrm{MAP}=\Lambda^{\overline{\mathrm{MAP}}}_{j,b}=\infty$}
($\forall\,j,b$). Whenever a leaf with corresponding label $\bmx$ has
been reached, the decoder distinguishes between two cases:

\subsubsection*{i) MAP Hypothesis Update} If
$d(\bmx)<\lambda^\mathrm{MAP}$, a new MAP hypothesis has been found.
First, all extrinsic metrics~$\Lambda^{\overline{\mathrm{MAP}}}_{j,b}$
for which $x_{j,b}=\overline{x^\mathrm{MAP}_{j,b}}$ are updated
according to
\begin{align*}
\Lambda^{\overline{\mathrm{MAP}}}_{j,b} \gets \IEmap{\lambda^\mathrm{MAP},L^A_{j,b},x_{j,b}^{\overline{\mathrm{MAP}}}}
\end{align*}
followed by the updates~$\lambda^\mathrm{MAP}\gets d(\bmx)$ and
$\bmx^\mathrm{MAP}\gets \bmx$. In other words, for each bit in the MAP
hypothesis that is changed in the update process, the metric
associated with the \emph{former} MAP hypothesis becomes the extrinsic
metric of the \emph{new} counter-hypothesis. 

\subsubsection*{ii) Extrinsic Metric Update}

If~$d(\bmx)>\lambda^\mathrm{MAP}$, only extrinsic metrics
corresponding to counter-hypotheses might be updated. For each
$j=1,2,\ldots,\MT$, $b=1,2,\ldots,Q$
with~$x_{j,b}=\overline{x^\mathrm{MAP}_{j,b}}$ and
$\IEmap{d(\bmx),L^A_{j,b},x_{j,b}^\mathrm{MAP}}<\Lambda^{\overline{\mathrm{MAP}}}_{j,b}$,
the SISO STS-SD performs the update
\begin{align} \label{eq:extrinsicmetricupdate}
 \Lambda^{\overline{\mathrm{MAP}}}_{j,b} \gets \IEmap{d(\bmx),L^A_{j,b},x_{j,b}^\mathrm{MAP}}.
\end{align}

\subsection{Extrinsic LLR Clipping}

In order to ensure that the extrinsic LLRs delivered by the algorithm
indeed satisfy \mbox{$\big|L^E_{j,b}\big|\leq L_\mathrm{max}$}
($\forall j,b$), the following update rule
\begin{align}  \label{eq:LLRclippingupdaterule}
\Lambda^{\overline{\mathrm{MAP}}}_{j,b}\gets
\min\Big\{\Lambda^{\overline{\mathrm{MAP}}}_{j,b},\lambda^\mathrm{MAP}+L_\mathrm{max}
\Big\}, \quad \forall j,b 
\end{align}
has to be applied after carrying out the steps in Case~i) of
the list administration procedure described
in~\secref{listadministration}.  We emphasize that
for~$L_\mathrm{max}=\infty$ the decoder attains max-log optimal SISO
performance, whereas for~$L_\mathrm{max}=0$, the hard-output MAP
solution~\eqref{eq:MAPsolution} is found.

\subsection{The Pruning Criterion}

Consider the node~$\bms^{(j)}$ on level~$j$ corresponding to the label
bits~$x_{i,b}$ (\mbox{$i=j,j+1,\ldots,\MT$}, \mbox{$b=1,2,\ldots,Q$}).
Assume that the subtree originating from this node and corresponding
to the label bits $x_{i,b}$ (\mbox{$i=1,2,\ldots,j-1$},
\mbox{$b=1,2,\ldots,Q$}) has not been expanded yet. The criterion for
pruning the node~$\bms^{(j)}$ along with its subtree is compiled from
two sets defined as follows:
\begin{itemize}
\item[1)] The bits in the partial label~$\bmx^{(j)}$ corresponding to
  the node~$\bms^{(j)}$ are compared with the corresponding bits in
  the label of the current MAP hypothesis.  All extrinsic
  metrics~$\Lambda^{\overline{\mathrm{MAP}}}_{j,b}$ with
  $x_{j,b}=\overline{x^\mathrm{MAP}_{j,b}}$ found in this comparison
  may be affected when searching the subtree originating from
  $\bms^{(j)}$. As the metric $d\big(\bmx^{(j)}\big)$ is an intrinsic
  metric, the extrinsic metrics
  $\Lambda^{\overline{\mathrm{MAP}}}_{j,b}$ need to be mapped to
  intrinsic metrics according
  to~\eqref{eq:extrinsictointrinsicmapping}. The resulting set of
  \emph{intrinsic} metrics, which may be affected by an update, is
  given by
\end{itemize}
\vspace{-0.1cm}
\begin{align*} 
  \setA_1\Big(\bmx^{(j)}\Big) = \Big\{
  \EImap{\Lambda^{\overline{\mathrm{MAP}}}_{i,b},L^A_{i,b},x_{i,b}^{\mathrm{MAP}}}  \,\Big|\,
   \big(i\geq j,\forall b\big) \\
   {\land}\:\Big(x_{i,b}=\overline{x^\mathrm{MAP}_{i,b}}\Big) \Big\}.
\end{align*}
\vspace{-0.3cm}
\begin{itemize}
\item[2)] The extrinsic metrics
  $\Lambda^{\overline{\mathrm{MAP}}}_{i,b}$ for $i=1,2,\ldots,j-1$,
  $b=1,2,\ldots,Q$ corresponding to the counter-hypotheses in the
  subtree of~$\bms^{(j)}$ may be affected as well. Correspondingly,
  we define
\end{itemize}
\vspace{-0.1cm}
\begin{align*}
  \setA_2\Big(\bmx^{(j)}\Big) & = \Big\{
  \EImap{\Lambda^{\overline{\mathrm{MAP}}}_{i,b},L^A_{i,b},x_{i,b}^\mathrm{MAP}}\,\Big|\,
  i<j,\forall b\Big\}.
\end{align*}

In summary, the intrinsic metrics which may be affected during the
search in the subtree emanating from node~$\bms^{(j)}$ are given by
\mbox{$\setA\big(\bmx^{(j)}\big) = \{a_l\} =
  \setA_1\big(\bmx^{(j)}\big)\cup\setA_2\big(\bmx^{(j)}\big)$}. The
node~$\bms^{(j)}$ along with its subtree is pruned if the
corresponding PD~$d\big(\bmx^{(j)}\big)$ satisfies the pruning
criterion
\begin{align*}
  d\Big(\bmx^{(j)}\Big) > \max_{a_l\in\setA\big(\bmx^{(j)}\big)} a_l.
\end{align*}
This pruning criterion ensures that a given node and the entire
subtree originating from that node are explored only if this could
lead to an update of either $\lambda^\mathrm{MAP}$ or of at least one
of the extrinsic metrics $\Lambda^{\overline{\mathrm{MAP}}}_{j,b}$.
Note that $\lambda^\mathrm{MAP}$ does not appear in the
set~$\setA\big(\bmx^{(j)}\big)$ as the update criteria given in
\secref{listadministration} ensure that $\lambda^\mathrm{MAP}$ is
always smaller than or equal to all intrinsic metrics associated with
the counter-hypotheses.


\section{Simulation Results} \label{simulationresults}

\begin{figure}[tb]
  \centering
  \includegraphics[width=0.95\columnwidth]{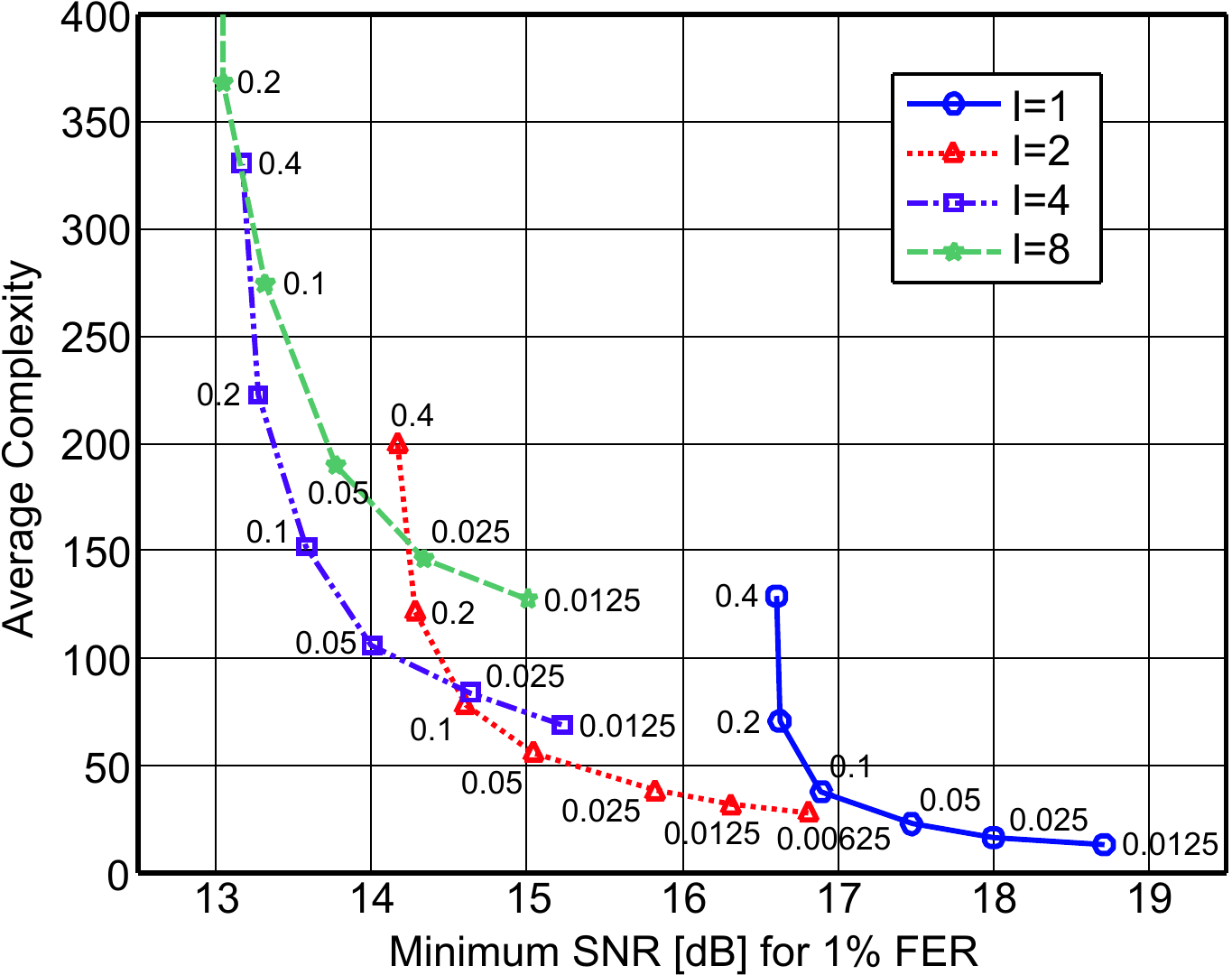}
\vspace{-0.15cm}
  \caption{Performance/complexity trade-off of the SISO STS-SD with
    sorted QR decomposition (SQRD) as described in~\cite{wuebben01}.
    The numbers next to the curves correspond to normalized LLR
    clipping levels and $\NIT$ denotes the number of iterations over
    the MIMO detector (and the channel decoder).}
  \label{fig:chclipiters}
\vspace{-0.45cm}
\end{figure}

\figref{fig:chclipiters} shows performance/complexity trade-off
curves\footnote{All simulation results are for a convolutionally
  encoded (rate $1/2$, generator polynomials [$133_o\,\,171_o$], and
  constraint length~7) MIMO-OFDM system with $\MT=\MR=4$,
  \mbox{16-QAM} symbol constellation with Gray labeling, 64~OFDM
  tones, a TGn type C channel model~\cite{TGN04}, and are based on a
  max-log BCJR channel decoder.  One frame consists of $1024$ randomly
  interleaved (across space and frequency) bits corresponding to one
  (spatial) OFDM symbol. The SNR is per receive antenna.} for the SISO
STS-SD described in Sections~\ref{SISOSD} and~\ref{SISOSTS}. The
numbers next to the SISO STS-SD trade-off curves correspond to
\emph{normalized} LLR clipping levels given by~$L_\mathrm{max}N_o$.
The average (over channel, noise, and data realizations) complexity
corresponds to the \emph{cumulative} tree-search complexity associated
with SISO detection over $\NIT$ iterations, where one iteration is
defined as using the MIMO detector (and the subsequent channel
decoder) once. The curve associated with $\NIT=1$ hence corresponds to
the soft-output STS-SD described in~\cite{jsac07}. Increasing the
number of iterations allows to reduce the SNR operating point (defined
as the minimum SNR required to achieve a frame error rate of 1\%) at
the cost of increased complexity. We can see that performance improves
significantly with increasing number of iterations. Note, however,
that for a fixed SNR operating point, the minimum complexity is not
necessarily achieved by maximizing the number of iterations~$\NIT$ as
the trade-off region is parametrized by the LLR clipping level
\emph{and} the number of iterations~$\NIT$.

\begin{figure}[t]
  \centering
  \includegraphics[width=0.95\columnwidth]{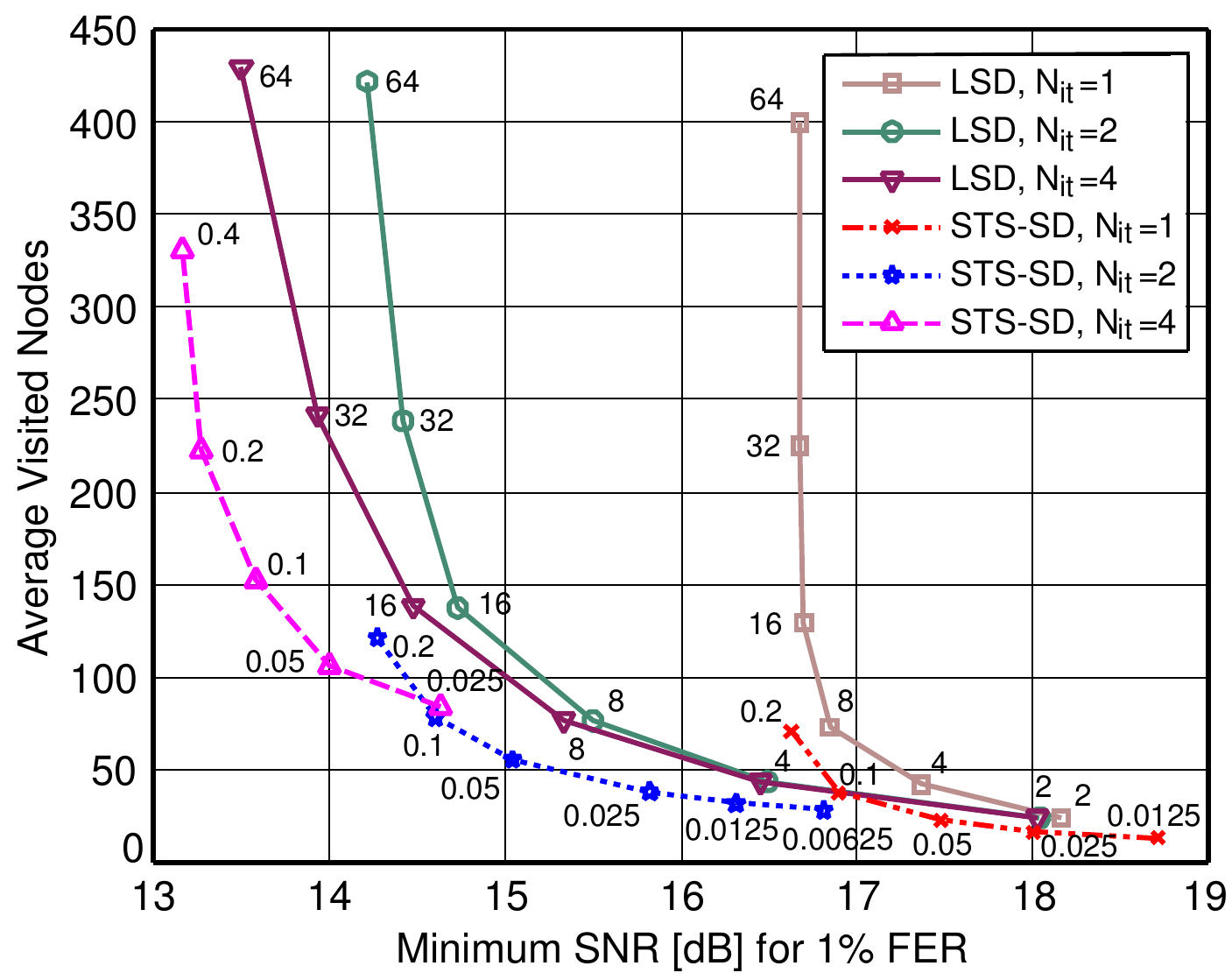}
\vspace{-0.15cm}
\caption{Performance/complexity (trade-off) comparison of the list
  sphere decoder (LSD)~\cite{HB03} and the SISO STS-SD (both using
  SQRD). The numbers next to the curves correspond to the list size
  for the LSD and to normalized LLR clipping levels for the SISO
  STS-SD.}
  \label{fig:chcomparison}
\vspace{-0.45cm}
\end{figure}

\figref{fig:chcomparison} compares the performance/complexity
trade-off achieved by the list sphere decoder (LSD)~\cite{HB03} to
that obtained through the SISO STS-SD. For the LSD we take the
complexity to equal the number of nodes visited when building the
initial candidate list. The (often significant) computational burden
incurred by the list administration of the LSD is neglected here. We
can draw the following conclusions from~\figref{fig:chcomparison}:
\begin{itemize}
\item[i)] The SISO STS-SD outperforms the LSD for all SNR values.
\item[ii)] The LSD requires relatively large list sizes and hence a
  large amount of memory to approach max-log optimum SISO performance.
  The underlying reason is that the LSD obtains extrinsic LLRs from a
  list that has been computed around the maximum-likelihood solution,
  i.e., in the absence of a priori information. In contrast, the SISO
  STS-SD requires memory mainly for the extrinsic metrics. The
  extrinsic LLRs are obtained through a search that is concentrated
  around the MAP solution. Therefore, the SISO STS-SD tends to require
  (often significantly) less memory than the LSD.
\end{itemize}

Besides the LSD, various other SISO detection algorithms for MIMO
systems have been developed, see
e.g.,\cite{tuechler02,Steingrimsson03,baero03,Hagenauer07}.
For~\cite{tuechler02,Steingrimsson03} issues indicating potentially
prohibitive computational complexity include the requirement for
multiple matrix inversions at symbol-vector rate. In contrast, the QR
decomposition required for sphere decoding has to be computed only
once per frame. The computational complexity of the list-sequential
(LISS) algorithm in~\cite{baero03,Hagenauer07} seems difficult to
relate to the complexity measure employed in this paper.  However, due
to the need for sorting of candidate vectors in a list and the
structural similarity of the LISS and the LSD algorithms, we expect
their computational complexity behavior to be comparable as well.

\bibliographystyle{IEEEtran} \bibliography{IEEEabrv,isit08}

\begin{thebibliography}{10}
\providecommand{\url}[1]{#1}
\csname url@rmstyle\endcsname
\providecommand{\newblock}{\relax}
\providecommand{\bibinfo}[2]{#2}
\providecommand\BIBentrySTDinterwordspacing{\spaceskip=0pt\relax}
\providecommand\BIBentryALTinterwordstretchfactor{4}
\providecommand\BIBentryALTinterwordspacing{\spaceskip=\fontdimen2\font plus
\BIBentryALTinterwordstretchfactor\fontdimen3\font minus
  \fontdimen4\font\relax}
\providecommand\BIBforeignlanguage[2]{{%
\expandafter\ifx\csname l@#1\endcsname\relax
\typeout{** WARNING: IEEEtran.bst: No hyphenation pattern has been}%
\typeout{** loaded for the language `#1'. Using the pattern for}%
\typeout{** the default language instead.}%
\else
\language=\csname l@#1\endcsname
\fi
#2}}

\bibitem{jsac07}
C.~Studer, A.~Burg, and H.~B\"olcskei, ``Soft-output sphere decoding:
  Algorithms and {VLSI} implementation,'' \emph{IEEE Journal on Selected Areas
  in Communications}, vol.~26, no.~2, pp. 290--300, Feb. 2008.

\bibitem{vikalo02a}
H.~Vikalo and B.~Hassibi, ``Modified {Fincke-Pohst} algorithm for
  low-complexity iterative decoding over multiple antenna channels,'' in
  \emph{Proc. of IEEE International Symposium on Information Theory (ISIT)},
  2002, p. 390.

\bibitem{jalden05}
J.~Jald\'en and B.~Ottersten, ``Parallel implementation of a soft output sphere
  decoder,'' in \emph{Proc. of Asilomar Conference on Signals, Systems and
  Computers}, Nov. 2005, pp. 581--585.

\bibitem{HB03}
B.~M. Hochwald and S.~ten Brink, ``Achieving near-capacity on a
  multiple-antenna channel,'' \emph{IEEE Trans. on Communications}, vol.~51,
  no.~3, pp. 389--399, Mar. 2003.

\bibitem{Steingrimsson03}
B.~Steingrimsson, T.~Luo, and K.~M. Wong, ``Soft quasi-maximum-likelihood
  detection for multiple-antenna wireless channels,'' \emph{IEEE Trans. on
  Signal Processing}, vol.~51, no.~11, pp. 2710--2719, Nov. 2003.

\bibitem{wang04}
R.Wang and G.~B. Giannakis, ``Approaching {MIMO} channel capacity with
  reduced-complexity soft sphere decoding,'' in \emph{Proc. of IEEE Wireless
  Communications and Networking Conference (WCNC)}, vol.~3, Mar. 2004, pp.
  1620--1625.

\bibitem{AEVZ02}
E.~Agrell, T.~Eriksson, A.~Vardy, and K.~Zeger, ``Closest point search in
  lattices,'' \emph{IEEE Trans. on Information Theory}, vol.~48, no.~8, pp.
  2201--2214, Aug. 2002.

\bibitem{burg05}
A.~Burg, M.~Borgmann, M.~Wenk, M.~Zellweger, W.~Fichtner, and H.~B\"olcskei,
  ``{VLSI} implementation of {MIMO} detection using the sphere decoding
  algorithm,'' \emph{{IEEE} Journal of Solid-State Circuits}, vol.~40, no.~7,
  pp. 1566--1577, July 2005.

\bibitem{Hagenauer96}
J.~Hagenauer, E.~Offer, and L.~Papke, ``Iterative decoding of binary block and
  convolutional codes,'' \emph{IEEE Trans. on Information Theory}, vol.~42,
  no.~2, pp. 429--445, Mar. 1996.

\bibitem{baero03}
S.~B\"aro, J.~Hagenauer, and M.~Witzke, ``Iterative detection of {MIMO}
  transmission using a list-sequential ({LISS}) detector,'' in \emph{Proc. of
  IEEE International Conference on Communications ({ICC})}, vol.~4, May 2003,
  pp. 2653--2657.

\bibitem{Yee05}
M.~S. Yee, ``{Max-Log-Map} sphere decoder,'' in \emph{Proc. of the IEEE
  International Conference on Acoustics, Speech, and Signal Processing
  (ICASSP)}, vol.~3, Mar. 2005, pp. 1013--1016.

\bibitem{TGN04}
V.~Erceg \emph{et~al.}, \emph{TGn channel models}, May 2004, {IEEE} 802.11
  document 03/940r4.

\bibitem{wuebben01}
D.~W\"ubben, R.~B\"ohnke, J.~Rinas, V.~K\"uhn, and K.-D. Kammeyer, ``Efficient
  algorithm for decoding layered space-time codes,'' \emph{IEE Electronics
  Letters}, vol.~37, no.~22, pp. 1348--1350, Oct. 2001.

\bibitem{tuechler02}
M.~T\"uchler, A.~C. Singer, and R.~Koetter, ``Minimum mean squared error
  equalization using a priori information,'' \emph{IEEE Trans. on Signal
  Processing}, vol.~50, no.~3, pp. 673--983, Mar. 2002.

\bibitem{Hagenauer07}
J.~Hagenauer and C.~Kuhn, ``The list-sequential ({LISS}) algorithm and its
  application,'' \emph{IEEE Trans. on Communications}, vol.~55, no.~5, pp.
  918--928, May 2007.

\end{thebibliography}

\end{document}